# Incorporating genuine prior information about between-study heterogeneity in random effects pairwise and network meta-analyses

Shijie Ren[1], Jeremy E. Oakley[2] and John W Stevens[3]

26 July 2017

[1] School of Health and Related Research, University of Sheffield, UK

s.ren@sheffield.ac.uk

[2] School of Mathematics and Statistics, University of Sheffield, UK

j.oakley@sheffield.ac.uk

[3] School of Health and Related Research, University of Sheffield, UK

j.w.stevens@sheffield.ac.uk

# Abstract

**Background**: Pairwise and network meta-analyses using fixed effect and random effects models are commonly applied to synthesise evidence from randomised controlled trials. The models differ in their assumptions and the interpretation of the results. The model choice depends on the objective of the analysis and knowledge of the included studies. Fixed effect models are often used because there are too few studies with which to estimate the between-study standard deviation from the data alone.

**Objectives**: The aim is to propose a framework for eliciting an informative prior distribution for the between-study standard deviation in a Bayesian random effects meta-analysis model to genuinely represent heterogeneity when data are sparse.

**Methods**: We developed an elicitation method using external information such as empirical evidence and experts' beliefs on the 'range' of treatment effects in order to infer the prior distribution for the between-study standard deviation. We also developed the method to be implemented in R.

**Results**: The three-stage elicitation approach allows uncertainty to be represented by a genuine prior distribution to avoid making misleading inferences. It is flexible to what judgments an expert can provide, and is applicable to all types of outcome measure for which a treatment effect can be constructed on an additive scale.

**Conclusions**: The choice between using a fixed effect or random effects meta-analysis model depends on the inferences required and not on the number of available studies. Our elicitation framework captures external evidence about heterogeneity and overcomes the often implausible assumption that studies are estimating the same treatment effect, thereby improving the quality of inferences in decision making.

# 1      Introduction

Evidence of clinical effectiveness can arise from multiple sources. Pairwise meta-analysis (MA) is an established statistical tool for estimating the relative efficacy of two interventions evaluated in randomised controlled trials. In the absence of head-to-head studies, indirect and mixed treatment comparison, also known as network meta-analysis (NMA) can be used to synthesise all available evidence and make simultaneous comparisons between treatments.

A pairwise MA and NMA can be conducted using a fixed effect or a random effects model. These models differ in their assumptions as well as in the interpretation of the treatment effects (1–4). The choice of which model to use depends on the objective of the analysis and knowledge of the included studies. In this paper, we investigate the circumstances when and rationale for using these two models in National Institute for Health and Care Excellence (NICE) single technology appraisals (STAs). We also propose how to overcome the problem of imprecise estimates of the heterogeneity parameter in the absence of sufficient sample data.

A fixed effect model would be appropriate if the objective is to determine whether the treatment had an effect in the observed studies (i.e. a conditional inference) and/or would be appropriate when it is believed that the true treatment effects in each study are the same. Heterogeneity is expected in MAs because they combine studies that have clinical and methodological heterogeneity (5). A random effects model would be preferred because it allows for heterogeneity in the treatment effects among the studies, and allows the results to be generalised beyond the studies included in the analysis. Nevertheless, fixed effect models are still commonly used even when heterogeneity is expected.

Parameters can be estimated and inferences can be made from either a frequentist or Bayesian perspective. The Bayesian approach provides more natural and useful inference, can incorporate external information and is deal for problems of decision making. There has been an increase in the use of Bayesian evidence synthesis in submissions to NICE, perhaps primarily because the evidence synthesis Technical Support Documents (TSDs) issued by the NICE Decision Support Unit (DSU) (6–11) advocate the Bayesian approach.

A random effects model requires an estimate of the between-study standard deviation (SD). In the case when the number of included studies is small, the estimate of the between-study SD will be highly imprecise and biased in a frequentist framework such as using DerSimonian and Laird estimate (1). Similarly, a Bayesian analysis of only limited data, using a standard vague /weakly informative prior distribution for the between-study SD will give implausible posterior distributions (12). A proper Bayesian analysis requires genuine specification of the prior distribution using external evidence, typically including experts' beliefs.

NICE TSD (7) suggests comparing goodness-of-fit of both fixed effect and random effects models using the deviance information criteria (DIC) (13). However, when the number of studies is small, it is likely that either model would at least provide an adequate fit to the data, specifically when the data are not sufficiently informative to learn about the between-study SD. Rather than goodness-of-fit, the issue is therefore how best to appropriately represent uncertainty about the treatment effect. When heterogeneity is expected, a fixed effect model is likely to be overconfident, and a random effects model with a vague prior is likely to be underconfident: a compromise between these two extremes is needed, which can be achieved with a more informative prior distribution.

Higgins et al (1996) (14) presented an example of a Bayesian meta-analysis of meta-analyses to create a predictive distribution for the between-study variance in the area of gastroenterology. Other authors have generated predicative distributions for the heterogeneity expected in future MAs in more general settings using data from the Cochrane Database of Systematic Reviews for a log odds ratio (OR) (15–17), and a standardised mean difference (18). Smith et al (1995) (19) constructed an informative prior distribution for the between-study variance using a gamma distribution by assuming that odds ratios between studies have roughly one order of magnitude spread, and that it is very unlikely that the variability in treatment effects between studies varies by two or more orders of magnitude. Spiegelhalter et al (2004) (20) suggested that a half-Normal distribution could be used as a prior distribution for the between-study SD and showed how to interpret the prior distribution. NICE TSD (8) also suggested using an informative half-Normal prior distribution with mean 0 and variance $0.32^2$, and provided an interpretation that it represents the belief that 95% of the study-specific ORs lie within a factor of 2 from the median OR for each comparison. Both of these half-Normal prior distributions are proposed for treatment effects measured by odds ratios. To the best of our knowledge, there has been little work on the formal elicitation of experts' beliefs for the between-study SD in random effects MA models.

To investigate the application of fixed effect and random effects models in submissions to NICE, we conducted a review of all of the NICE STAs completed up to 31st October 2016. Results of the review are presented in Section 2. In Section 3, we propose novel methods to construct an informative prior distribution for the between-study SD utilising external information for all common types of outcome measures. Examples of re-analysing two STAs using the proposed elicitation framework are given in Section 4.

## 2    A review of NICE STAs

239 NICE STAs were completed between September 2005 (when the STA process was introduced) and 31st October 2016. After assessment by SR, a final set of 183 STAs was identified for review. Figure 1 presents a flow chart of identification, inclusion and exclusion of STAs. We have only reviewed the original companies' submissions, and not considered additional analyses that may have occurred during the appraisal process.

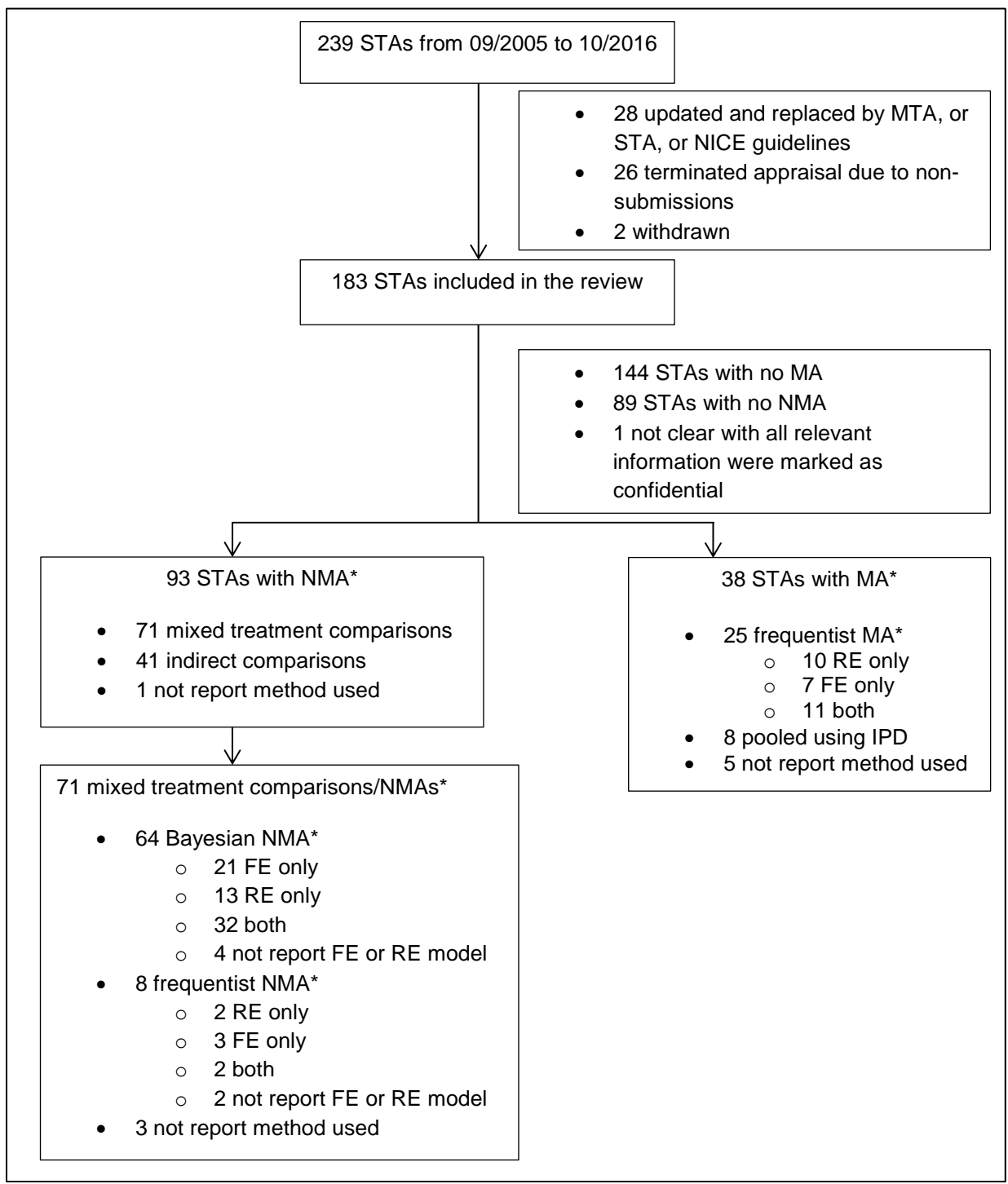

Figure 1: Flow chart showing the identification, inclusion and exclusion of reviews. Abbreviations: STA for single technology appraisal, MTA for multiple technology appraisal, RE for random effects, FE for fixed effect, MA for meta-analysis and NMA for network meta-analysis, IPD for individual patient-level data. *: multiple analyses and analyses for multiple outcomes may have been conducted in one submission.

38 STA submissions used standard pairwise MAs with a single approach being applied within each submission: 25 (66%) used a frequentist approach to estimate parameters and make inferences; 8 (21%) pooled individual patient-level data across studies; and in 5 cases (13%) it was unclear which method was used. 93 STA submissions included NMAs (multiple approaches may have been used in one submission). 71 (76%) of these used either a Bayesian or a frequentist NMA, 41 (44%) used Bucher indirect comparisons (21) and 1 (1%) didn't report the method that was used.

We extracted the rationale for using fixed effect and random effects model for both pairwise MAs and NMAs. The findings of the review are presented in Table 1 and are summarised as follows:

- All of the submissions that performed pairwise MAs used a frequentist approach, and the majority of the submissions that performed NMAs used a Bayesian approach (90%).

- 71% of the submissions that performed fixed effect pairwise MAs did not provide a justification for the model choice. For the submissions that performed random effects MAs, 60% of them gave no justification for the model choice. Fewer submissions used NMAs provided no justification for the model choice, 25% and 27% for fixed effect and random effects model, respectively.

- The most frequently stated reason for the use of a fixed effect model was that there were too few studies to conduct a random effects model.

- In some cases, where heterogeneity was noted, there was an acknowledgement that a random effects model would be appropriate, although it was not used when there were only few studies.

- Among the pairwise MAs that used a frequentist approach, the choice of fixed effect or random effects model was typically assessed using the Q-statistic/$I^2$-statistic.

- When Bayesian fixed effect and random effects models were both used in a submission, the most popular method for choosing the final model was comparing the DIC statistic for the two models (62%).

- Providing either a fixed effect or random effects model within a sensitivity analysis was observed in both pairwise MAs (9%) and NMAs (21%) in the case where both models were used.

- Four submissions performed sensitivity analyses using different prior distributions for the between-study SD. TA288 (22) considered the possibility of using alternative data sources to inform the prior distribution but concluded that no suitable sources were available. TA341 (23) used a prior distribution informed using predictive distributions proposed by Turner el al (2012) (16). TA173 (24) used a half-Normal prior based on a re-analysis of the data from a previous systematic review. TA343 (25) used a half-Normal prior distribution suggested by NICE TSD (8).

| Method used (number of submissions) | | Justification | N (%) |
|---|---|---|---|
| Pairwise meta-analysis (38*) | Fixed effect model only (7) | No justification | 5 (71%) |
| | | Check heterogeneity using test statistic | 2 (29%) |
| | Random effects model only (10) | No justification | 6 (60%) |
| | | Allow for heterogeneity | 3 (30%) |
| | | Check heterogeneity using test statistic | 1 (10%) |
| | Both models (11**) | Not clear which model was a base case | 5 (45%) |
| | | Check heterogeneity using test statistic | 4 (36%) |
| | | One model as sensitivity analysis | 1 (9%) |
| | | Checking inclusion criteria | 1 (9%) |
| | Pooling using individual patient-level data (8) | | |
| | Unclear (5) | | |
| Network meta-analysis (71*) | Fixed effect model only (24) | Insufficient data | 17 (71%) |
| | | No justification | 6 (25%) |
| | | Check heterogeneity using test statistic | 1 (4%) |
| | Random effects model only (15) | Allow for heterogeneity | 4 (27%) |
| | | No justification | 4 (27%) |
| | | Sufficient data | 2 (13%) |
| | | Check heterogeneity using test statistic | 1 (7%) |
| | | Same model as a previous study | 1 (7%) |
| | | Count for correlations | 1 (7%) |
| | | Count for multi-arms | 1 (7%) |
| | | Unclear | 1 (7%) |
| | Both models (34**) | Based on deviance information criteria | 21 (62%) |
| | | One model as sensitivity analysis | 7 (21%) |
| | | Final model fixed effect because of insufficient data | 4 (12%) |
| | | Not clear which model was a base case | 3 (9%) |
| | | Compare the credible intervals | 1 (3%) |
| | | Presence of closed loops | 1 (3%) |
| | Unclear (2) | | |

Table 1: Justifications of model choice in submissions. *: multiple analyses and analyses for multiple outcomes may have been conducted in one submission. **: multiple reasons for model choice may have been used in one analysis.

Overall, we found that the most frequently stated reason for the use of a fixed effect model was that there were too few studies to conduct a random effects model, but not that there was unlikely to be heterogeneity or that a conditional inference was of interest. This showed that there is a need for more guidance on properly accounting for heterogeneity when the number of included studies is small.

We now present a framework for constructing prior distributions for the heterogeneity parameter using external information such as empirical evidence and experts' beliefs.

## 3 General elicitation framework

For simplicity, we suppose that there is one female expert and the elicitation is conducted by a male facilitator. We do not consider issues such as the selection and training of experts, motivation and how to elicit a prior distribution from multiple experts, which are covered elsewhere (26–28). The general elicitation framework proposed in this section is for performing a pairwise MA. An extension to the approach for use in NMAs is discussed later.

We envisage that the elicitation will take place after specification of the decision problem and completion of the systematic literature review, and that this finds that there are few studies that satisfy the inclusion/exclusion criteria for the MA. The expert making the judgments could be a clinician, or an analyst who conducts the MA. She will be given the information on the decision problem, including population, intervention, control and outcome, summary of the included studies, and encouraged to think about any potential treatment effect modifiers.

Suppose that there are $S$ studies in the included in the MA, and that the treatment effect in study $i$ is denoted by $\delta_i$, for $i = 1 \ldots S$, expressed on some appropriate additive scale. The expert is required to make judgements about the likely variability in $\delta_1, \ldots, \delta_S$ between studies. For any two studies $i$ and $j$, one could make judgements about the relative treatment effect $\delta_i/\delta_j$, i.e. 'the treatment effect in one study could be $x$ times that of the treatment effect in another'. Alternatively, one could consider the difference in treatment effects $\delta_i - \delta_j$, i.e., 'the treatment effect in one study could exceed that in another study by $x$ units'. In this paper, we consider the former case only, building on the discussion and analysis by previous authors (19,20). In the latter case, elicitation methods for variances discussed in (29) could be considered.

We assume that $\delta_1, \ldots, \delta_S \sim N(d, \tau^2)$, where $d$ is the average treatment effect in a population of treatment effects and $\tau^2$ is the between-study variance, which represents the heterogeneity in treatment effects between studies. This is the standard model when $\delta_i$ is a log OR, log hazard ratio, or mean difference (7). We define $R$ to be the ratio of the 97.5th percentile to the 2.5th percentile of treatment effects on the natural scale from a population of treatment effects. When the additive treatment effect that is estimated in the MA is on the log scale, we propose to elicit the treatment effect on the natural scale. For example, if $\delta_i$ is a log OR, then we propose using $OR_i$ in the elicitation, which is $\exp(\delta_i)$, and $R = OR_{97.5}/OR_{2.5}$ is the ratio of the 97.5th percentile to the 2.5th percentile of ORs in a population of treatment effects, roughly representing the 'range' of ORs. Noting that $\log R = \delta_{0.975} - \delta_{0.025}$, we link $R$ to $\tau$ via

$$\delta_{97.5} - \delta_{2.5} = 2 \times 1.96\tau = 3.92\tau$$
$$\Rightarrow \log R = 3.92\tau$$
$$\Rightarrow \tau = \frac{\log(R)}{3.92} \qquad (1)$$

We propose asking the expert to make judgements about $R$, from which judgements about $\tau$ can be inferred using equation (1). However, given the somewhat abstract nature of $R$, we suggest providing the less formal definition to the expert: she is asked to consider the ratio of the largest to the smallest treatment effect on the natural scale that could arise over a set of studies (though the expert should be told that 'largest' and 'smallest' will be interpreted as 97.5th and 2.5th percentiles).

## 3.1 A three-stage procedure for eliciting the prior distribution for the between-study SD

In some cases, even with adequate training, the expert may find it difficult to make the judgements about $R$ that are necessary to obtain a distribution for $\tau$. Consequently, we propose a three-stage procedure depending on the judgements that the expert is able to make. We firstly present this procedure when the treatment effect is a log OR, and then discuss modifications for treatment effects reported on different scales. Code for implementing our method using R (31) is available online. Instructions are given in Appendix 3.

### 3.1.1 Stage 1: confirmation of the need for a random effects model

The fixed effect model is a special case of the random effects model, corresponding to the judgement that $P(R = 1) = 1$, i.e. the expert is certain that the largest OR is the same as the smallest OR in a set of studies. The expert is asked to either rule out or accept this case:

- "Can you be certain that the treatment effects across the studies will be identical, ignoring within-study sampling variability?"

If she is certain that this will be the case, then a fixed effect model should be used with appropriate justification provided. Otherwise, we proceed to Stage 2.

### 3.1.2 Stage 2: consideration of an upper bound for $R$

If a random effects model is deemed to be appropriate, the expert is then asked if she is able to provide an upper bound for $R$. She is asked:

- "Let $R$ be the ratio of the largest to the smallest OR. Are you able to judge a maximum plausible value for $R$? Denoting this limit by $R_{max}$, this means that you would think values of $R$ above $R_{max}$ are too implausible to be contemplated."

If the expert's answer for $R_{max}$ is, for example, 10, this means that she believes that the OR in one study could be no more than 10 times that of the OR in another, i.e. one order of magnitude. If the expert is not able to provide a value $R_{max}$, then we recommend using the prior distribution proposed by other authors (15–17). For example, Turner et al (2012) (16) proposed a prior distribution for the between-study variance in a general setting with the treatment effect measured by a log OR,

$$\log \tau^2 \sim N(-2.56, 1.74^2). \quad (2)$$

If she is able to provide $R_{max}$, then we proceed to Stage 3. Note that we do not propose asking for a lower limit for $R$ because we think experts would typically not want to rule out the case $R = 1$ as impossible. The expert could also provide a lower limit $R_{min}$, if she wished, with $R_{min}$ replacing the lower limit of 1 in the following.

### 3.1.3 Stage 3: consideration of a full distribution for $R$

We now ask if the expert judges some values in the range $[1, R_{max}]$ to be more likely than others, and if she is able to express her beliefs using the roulette elicitation method (30). If she is not able to make such judgements, then we propose using prior distributions proposed by (15–17), but now truncated to $\left[0, \left(\frac{\log(R_{max})}{3.92}\right)^2\right]$.

If she is able to continue with the roulette method, then the range $[1, R_{max}]$ is divided into a number of equal-width 'bins'. The expert is asked to specify her probability of $R$ lying in a particular bin by placing 'chips' in that bin, with the proportion of chips allocated representing her probability. The number of chips given to the expert is specified by the facilitator. For example, if in total 20 chips are used, then each chip represents a probability of 0.05. An illustration is given in Figure 2a. Here, the expert has chosen $R_{max} = 10$. By placing 5 chips out of 20 in the bin [2, 3], she has expressed a judgement that $P(2 < R \leq 3) = \frac{5}{20}$.

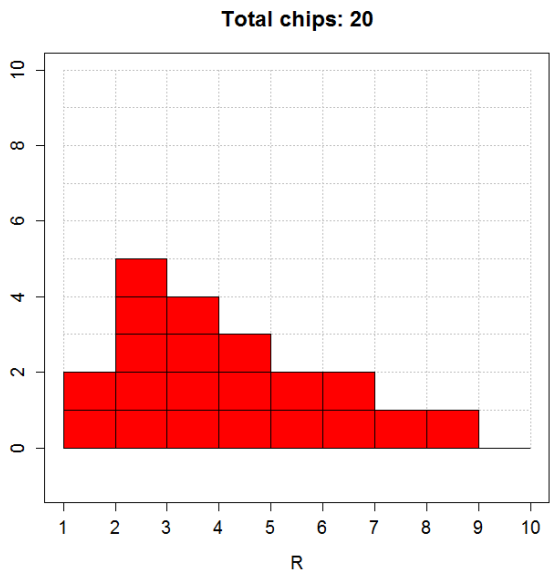
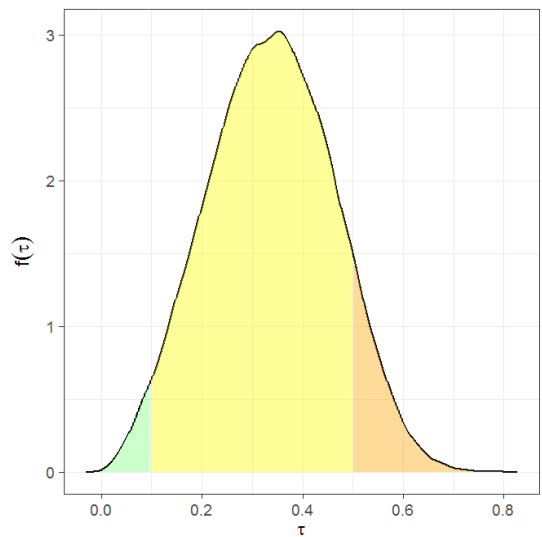

Figure 2a                                    Figure 2b

Figure 2: (a) Eliciting beliefs about $R$ with the roulette method. (b) The implied distribution of $\tau$, following the elicited judgements about $R$ shown in the left hand side Figure 2a. The probabilities of 'low', 'moderate' and 'high' heterogeneity are in green, yellow and orange, respectively (with negligible probability of 'extreme' heterogeneity).

We suggest fitting either a gamma or lognormal distribution to the elicited probabilities by choosing the distribution parameters to minimise the sum of squares between the elicited and fitted cumulative probabilities. The R code provided will identify the best fitting distribution out of the gamma and lognormal, although there is unlikely to be much difference in the fitted distributions in most practical situations. Given that $R$ has a lower limit of 1, the package will fit a gamma or lognormal distribution to $R - 1$. Hence, using a lognormal distribution for example, we will have a prior for $\tau$ specified via

$$\log(R - 1) \sim N(m, v),$$

$$\tau = \frac{\log R}{3.92},$$

where $m$ and $v$ are the mean and variance for the elicited lognormal distribution.

## 3.2 Feedback

We propose providing feedback to the expert about the implied distribution of $\tau$, regardless of which stage in the above procedure. Spiegelhalter et al (2004) (20) suggested that values of $\tau$ between 0.1 and 0.5 are considered as reasonable heterogeneity in many contexts (what we describe as moderate), from 0.5 to 1.0 as fairly high heterogeneity (what we describe as high) and above 1.0 fairly extreme heterogeneity (what we describe as extreme). The probability of $\tau$ in the range of below 0.1, (0.1, 0.5), (0.5, 1.0) and above 1.0 will be provided to the expert. The distribution can be displayed using a kernel density estimate or histogram of a large sample of randomly generated values of $\tau$. An illustration using `SHELF` package is given in Figure 2b. The probabilities of 'low', 'moderate' and 'high' heterogeneity are approximately 0.03, 0.87 and 0.1 respectively (with negligible probability of 'extreme' heterogeneity) given the elicited judgements about $R$ in Figure 2a.

## 3.3 Other type of outcome measures

Other scale-free outcome measures include hazard ratio, relative risk and ratio of means for continuous outcomes (31,32). The three-stage procedure could be used in these cases, although it is less clear that the prior distributions proposed by previous authors (15–17) would be appropriate because the distributions were derived based on empirical evidence of heterogeneity in ORs in MAs. It is likely that an elicitation exercise of considering a full distribution for the ratio of treatment effects $R$, for example the ratio of the largest to the smallest hazard ratio among the studies, would be required in these cases.

When the outcome measure is continuous or ordered categorical with the MA model using the identity or probit link functions, the expert may find it difficult to express beliefs about the 'range' of treatment effects because the continuous measurement is not unit-free and the probit scale is difficult to interpret directly. We propose using the method described in section 3.1 with the following modification:

1. Dichotomise the response using some appropriate cut-off $c$, to define a new treatment effect $\delta_i$ on the OR scale.
2. Considering ORs for the dichotomised response, use the three-stage procedure to elicit a prior distribution for $\tau$, the variability in log ORs in a population of studies.
3. Given a prior distribution for $\tau$, convert this to a prior distribution for the between-study SD $\tilde{\tau}^2$ on the original scale (i.e. probit or continuous) via

$$\tilde{\tau} = \omega\tau,$$

with $\omega = \frac{\sqrt{3}}{\pi}$ for the probit scale, and $\omega = \sigma\frac{\sqrt{3}}{\pi}$ for the continuous scale, where $\sigma$ is an estimate of an individual level standard deviation. The estimate could be a summary measure of the SDs in the included studies, pooled from included studies, or obtained from a single representative study.

Details of the derivation can be found in Appendix 1.

### 3.4 Network meta-analysis

NMAs typically assume an homogeneous variance model (7,14,33). A similar elicitation method as described above can be used to elicit the common heterogeneity parameter in an NMA. We suggest asking the expert for the 'range' of treatment effects $R$ for a pairwise comparison based on the one that the expert is most comfortable about in expressing her beliefs. When giving feedback to the expert on the probability that the heterogeneity would be low, moderate, high and extremely high, we could ask the expert whether she would agree with these elicited probabilities for other pairwise comparisons in the network.

## 4 Examples: reanalysis of two STAs

We re-analyse the data from two NICE STAs (TA163 (34) and TA336 (35)) to demonstrate the use of our proposed method. BUGS code incorporating the different prior specifications is provided in Appendix 2.

TA163 (34) was a technology appraisal of infliximab for treating acute exacerbations in adults with severely active ulcerative colitis. Data were available from 4 studies of 3 treatments (placebo, infliximab and ciclosporin) (Figure 3). The outcome measure was the colectomy rate at 3 months. A fixed effect model was used in the original submission. Table 2 presents results from a Bayesian NMA using a fixed effect model, a random effects model with a vague prior distribution uniform [0, 5], as used in Dias et al, (2013) (7) and three alternative informative prior distributions: the prior distribution in equation (2), both untruncated and truncated so that $R_{max} \leq 10$, and an elicited prior distribution using proposed method in Section 3. The elicited judgements were those of the author's SR. Results are presented as the median with 95% credible and prediction intervals based on 40000 iterations from the Markov chain after a burn-in of 60000 iterations using the software OpenBUGS (36).

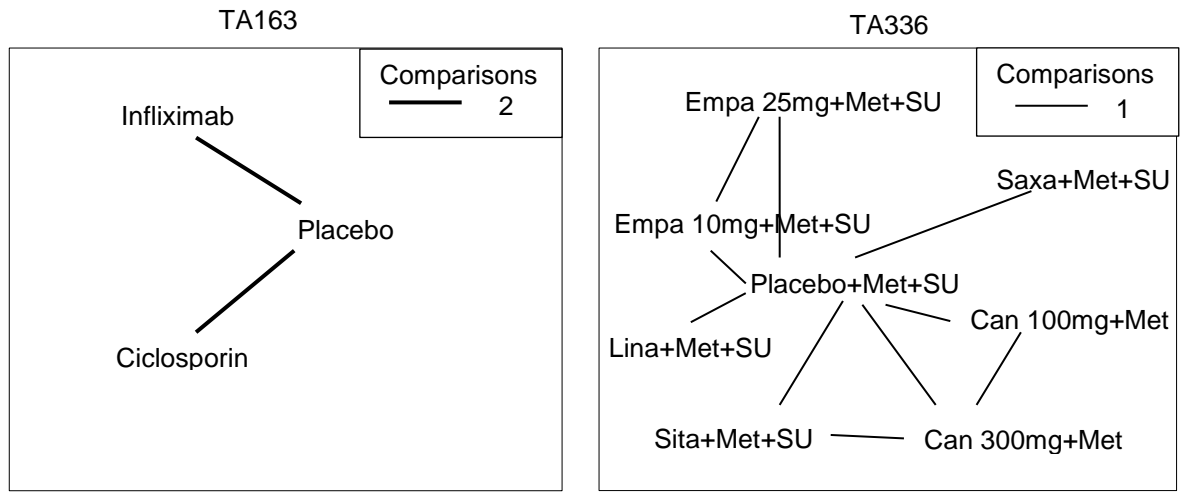

Figure 3: Network diagram for TA163 and TA336 used in the example. The thickness of the line represents the number of times pairs of treatment have been compared in studies. Abbreviations: Empa for empagliflozin, Lina for linagliptin, Sita for sitagliptin, Saxa for saxagliptin, Can for canagliflozin, Met for metformin and SU for sulphonylurea.

As expected, the DIC statistics for the four models were fairly similar: 34.72, 33.44, 34.70, 35.19 and 34.60 and did not provide strong support for any one model over the others. The fixed effect model showed that there was evidence that ciclosporin reduced the colectomy rate at 3 months relative to placebo in the studies included in the NMA, whereas there was insufficient evidence to conclude that infliximab had an effect relative to placebo in the included studies. As expected, the results of the random effects models demonstrated the sensitivity of the results to the different prior beliefs about the heterogeneity parameter. The uniform [0, 5] prior for the heterogeneity parameter was not 'updated' appreciably by the data (Figure 4) and gave very different results compared to the fixed effect model (Table 2). There was a large posterior probability, 0.87, that heterogeneity was extremely high, equivalent to saying that the probability that the OR in a study could be 50 or more times that of the OR in another was 0.87 (The interpretation of the heterogeneity parameter can be found in Appendix 1). This is unlikely to be plausible and the results using this prior distribution would not lead to reasonable posterior beliefs.

Results using empirical evidence as prior distribution and elicited prior distributions for the heterogeneity were much less uncertain than those produced using the uniform prior distribution but, as expected, differed depending on which prior distribution was used (Table 2). Using the untruncated lognormal prior, there was a small posterior probability that heterogeneity was extremely high, 0.08. The truncated lognormal and elicited prior distributions for the heterogeneity parameter both provided zero posterior probability of extreme values for the between-study SD. The truncating eliminated the possibility of extreme heterogeneity, i.e. the largest OR in one study could be no more than 10 times the OR in another study. The elicited prior distribution can be found in Appendix 3, which resulted in the probability of heterogeneity being low, moderate and high as 0.01, 0.85, and 0.14, respectively. The analyses using informative prior distributions for the heterogeneity parameter all suggested that ciclosporin reduced the colectomy rate at 3 months compared to placebo based on both the credible and prediction intervals, but the effect of infliximab versus placebo was inclusive. The credible and predictive intervals in the analyses using empirical evidence and elicited prior distributions were wider than the fixed effect interval because of the extra uncertainty but more plausible than analyses based on the fixed effect model and random effects model with a uniform prior distribution.

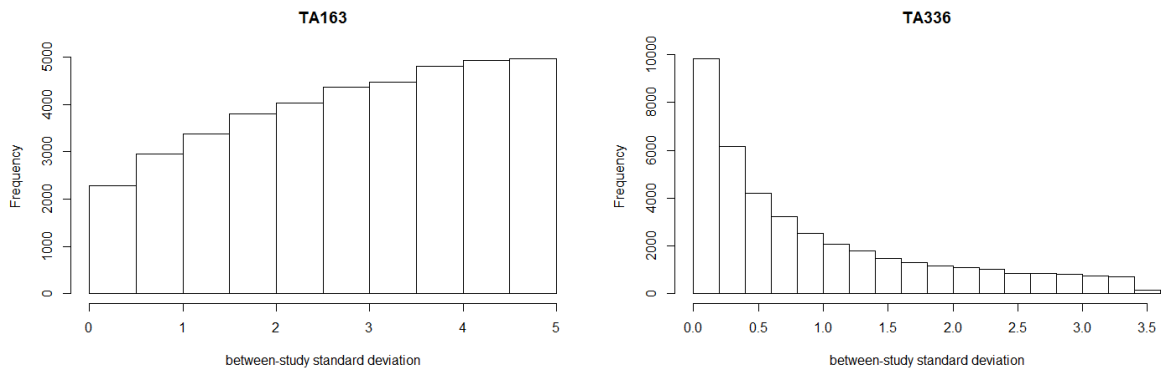

Figure 4: Posterior histogram plot of the between-study standard deviation using prior distribution as uniform [0,5].

TA336 (35) was a technology appraisal of empagliflozin for the treatment of Type 2 diabetes mellitus (T2DM). Data were available from 6 studies of 8 treatments in combination with metformin (Met) or metformin and sulphonylurea (Met+SU) (placebo, 10 and 25mg empagaliflozin, linagliptin, sitagliptin, saxagliptin, and 100 and 300mg canagliflozin). The outcome measure re-analysed here is the change from baseline in body weight for the third line treatment of T2DM at 24 weeks. A fixed effect model was used in the original submission.

Table 2 presents the results of 10mg empagliflozin + Met + SU versus placebo + Met + SU and linagliptin + Met + SU as an illustration. The DIC statistics for the four models were again similar: 3.82, 4.94, 4.61, 5.01 and 4.65. The fixed effect model showed that 10 mg empagliflozin +Met + SU reduced the change from baseline in body weight compared to placebo + Met + SU and linagliptin + Met + SU. When using the uniform [0, 5] prior distribution for the heterogeneity parameter there was more 'updating' in this case (Figure 4), but still a large posterior probability, 0.35, that the heterogeneity was extremely large. There was a small probability, 0.02, that the heterogeneity was extremely large when untrucated lognormal was used as the prior. The truncating eliminated the possibility of extreme heterogeneity in treatment effects between studies. The elicited prior can be found in Appendix 3, which resulted in the probability of heterogeneity being low, moderate and high as 0.06, 0.88, and 0.06, respectively. The analyses using informative prior distributions for the heterogeneity parameter all suggested that empagliflzin 10 mg is associated with beneficial treatment effect compared to placebo or linagliptin (all in combination with Met and SU) based on the credible and prediction intervals, with the exception of the prediction interval using untrucated lognormal prior.

| Example 1: TA163. Colectomy rate at 3 months: treatment effect on log odds ratio scale | OR, median (95% CrI) ciclosporin vs. placebo | OR, median (95% CrI) infliximab vs. placebo | $P_L$ | $P_M$ | $P_H$ | $P_{EH}$ |
|---|---|---|---|---|---|---|
| FE | 0.13 (0.03, 0.44) | 0.72 (0.18, 2.70) | 0 | 0 | 0 | 0 |
| RE with $\tau_{OR} \sim uniform\,[0,5]$ | 0.02 (0, 1.46) **0.03 (0, 33.02)** | 0.70 (0.01, 84.59) **0.69 (0, 2498.82)** | 0.01 | 0.05 | 0.07 | 0.87 |
| RE with $\tau_{OR}^2 \sim \log normal\,(0.256, 1.74^2)$ | 0.11 (0.01, 0.48) **0.12 (0.01, 0.62)** | 0.71 (0.14, 3.25) **0.71 (0.10, 4.83)** | 0.11 | 0.62 | 0.18 | 0.08 |
| RE with $\tau_{OR}^2 \sim$ truncated $\log normal\,(0.256, 1.74^2)\,I(0, 0.345)$ | 0.12 (0.03, 0.48) **0.12 (0.03, 0.54)** | 0.69 (0.17, 2.77) **0.69 (0.15, 3.14)** | 0.15 | 0.78 | 0.07 | 0 |
| RE with $(R_{OR} - 1) \sim gamma(2.62, 0.721)$ and $\tau_{OR} = \log(R_{OR} + 1)/3.92$ | 0.12 (0.03, 0.47) **0.12 (0.02, 0.56)** | 0.71 (0.17, 2.97) **0.71 (0.14, 3.69)** | 0.01 | 0.85 | 0.14 | 0 |
| | | | | | | |
| Example 2: TA336. Change from baseline in body weight at 24 weeks, kg: treatment effect on mean difference scale | MD, median (95% CrI) Empa 10mg+Met+SU vs. placebo+Met+SU | MD, median (95% CrI) Empa 10mg+Met+SU vs. linagliptin+Met+SU | $P_L$ | $P_M$ | $P_H$ | $P_{EH}$ |
| FE | -1.77 (-2.18, -1.35) | -2.10 (-2.64, -1.54) | 0 | 0 | 0 | 0 |
| RE with $\tau_{MD} \sim uniform\,[0,5]$ | -1.76 (-6.10, 2.70) **-1.76 (-7.70, 4.38)** | -2.08 (-8.12, 4.08) **-2.08 (-10.75, 6.55)** | 0.14 | 0.32 | 0.19 | 0.35 |
| RE with $\tau_{OR}^2 \sim \log normal\,(0.256, 1.74^2)$ and $\tau_{MD} = 2.61 \times \tau_{OR}/1.81$ | -1.77 (-2.88, -0.63) **-1.77 (-3.27, -0.18)** | -2.10 (-3.65, -0.51) **-2.10 (-4.22, 0.13)** | 0.18 | 0.70 | 0.10 | 0.02 |
| RE with $\tau_{OR}^2 \sim$ truncated $\log normal\,(0.256, 1.74^2)\,I(0, 0.345)$ and $\tau_{MD} = 2.61 \times \tau_{OR}/1.81$ | -1.77 (-2.62, -0.93) **-1.77 (-2.95, -0.63)** | -2.10 (-3.30, -0.93) **-2.10 (-3.74, -0.46)** | 0.21 | 0.75 | 0.04 | 0 |
| RE with $(R_{OR} - 1) \sim gamma\,(1.94, 0.741)$ and $\tau_{MD} = 2.61 \times \log(R_{OR} + 1)/(3.92 \times 1.81)$ | -1.78 (-2.76, -0.80) **-1.77 (-3.11, -0.45)** | -2.10 (-3.47, -0.72) **-2.10 (-3.98, -0.23)** | 0.08 | 0.88 | 0.03 | 0 |

Table 2: Comparison of results obtained from fixed effect and random effects models. $P_L$, $P_M$, $P_H$ and $P_{EH}$ denote the probability that heterogeneity being low, moderate, high and extremely high. Truncated log normal distribution has upper bound 0.345 representing that the 'range' of odds ratios between studies cannot exceed 10. Results in bold are the predictive distributions of the effects of treatments in a new study. Abbreviations: FE for fixed effect, RE for random effects, OR for odds ratio, CrI for credible interval, MD for mean difference, Empa for empagliflozin, Met for metformin and SU for sulphonylurea, vs for versus.

## 5   Discussion

Our review of NICE STAs showed that 17 (71%) out of 24 fixed effect NMAs were chosen on the basis that there were too few studies with which to estimate the heterogeneity parameter, but not that there was unlikely to be heterogeneity or that a conditional inference was of interest. A consequence of this is that decision uncertainty may be underestimated. The choice between using a fixed effect or random effects MA model depends on the inferences required and not on the number of studies. While a fixed effect model is informative in assessing whether treatments were effective in the observed studies, when we expect heterogeneity between studies and want to make unconditional inferences and predictions about the treatment effect in a new study, a random effects model should be used.

When heterogeneity is expected, the simple framework we have proposed overcomes the inappropriate assumption behind the use of a fixed effect model. We argue that in the absence of sufficient sample data, a minimum requirement should be to exclude extreme and implausible values from the prior distribution and the common choice of the prior distribution such as uniform [0, 5] or [0, 2] should not be used. We have shown in the examples that the use of a uniform prior distribution when data are sparse would result in implausible estimate for the heterogeneity parameter and unreasonable results for the treatment effect.

Our proposed elicitation framework is flexible with the amount of information an expert is able to provide. The minimum information required from the expert is the maximum possible value of the 'range' of treatment effects on the natural scale. For example, if the additive treatment effect is a log OR, then the expert is asked whether the OR in one study could be $x$ times that of the OR in another and what the maximum plausible value of $x$ could be. If the expert is not able to provide any judgments on the 'range' of treatment effects, then empirical evidence such as a prior distribution proposed for the heterogeneity expected in future MAs (15–18) could be considered. When the expert is able to provide only the maximum value of the 'range' of treatment effects, the prior distributions proposed by other authors (15–18) should be truncated accordingly before being used in the analysis. If the expert is able to provide complete probability judgments, then our proposed framework could facilitate the elicitation exercise.

In terms of presenting the results, we propose reporting the prior and posterior probabilities of heterogeneity being low, moderate, high and extremely high rather than simply as the point estimate and the credible interval, thereby presenting more information about the consequences of the chosen prior distribution. We also advocate the use of prediction intervals for the treatment effects as proposed by other authors (2,3,8). Prediction intervals provide a summary of the treatment effect expected in a new study which is more relevant to decision making.

In summary, it is important to incorporate genuine prior information about the heterogeneity parameter in a random effects pairwise MA/NMA in the absence of sufficient sample data with which to estimate it. Eliciting probability judgments from experts is not straightforward but is important if the aim is to genuinely represent uncertainty in a justifiable and transparent manner to properly inform decision making. Our proposed elicitation framework utilises external information such as empirical evidence and experts' beliefs, in which the minimum requirement from the expert is the maximum value of the 'range' of treatment effects. The method also is applicable to all types of outcome measure for which a treatment effect can be constructed on an additive scale.

# Appendix 1

## 1  Elicitation method for continuous outcome measures

For a continuous outcome measure, let $X_{ij}$ denote the sample mean in study $i$ on treatment arm $j$, with $j = 1$ the control arm and $j = 2$ the experimental treatment arm. Suppose the sample means have the distributions $X_{i1} \sim N(\mu_i, \sigma^2)$ and $X_{i2} \sim N(\mu_i + \beta_i, \sigma^2)$. Note that in the following, we assume the variances $\sigma^2$ are equal across arms and studies. The treatment effect, mean difference (MD) $\beta_i$, is on the original scale. A standardised mean difference (SMD), $\phi_i = \frac{\beta_i}{\sigma^2}$, may be used in meta-analysis if the included studies used different scales.

We further assume that the study-specific treatment effects are normally distributed: $\beta_1, \dots, \beta_S \sim N(d_{MD}, \tau_{MD}^2)$, or $\phi_1, \dots, \phi_S \sim N(d_{SMD}, \tau_{SMD}^2)$ depending on the scale used in each study. We suppose that the expert again prefers to consider variability in treatment effects via ratios of treatment effects, and we now consider a modification of the three-stage approach in Section 3.1.

If we can relate the treatment effects $\beta_i$ or $\phi_i$ to an odds ratio (OR) $\delta_i$, we could derive a distribution for $\tau_{MD}$ (the variability in mean differences (MDs) in a population of treatment effects) or $\tau_{SMD}$ (the variability in standardised mean differences (SMDs) in a population of treatment effects) via a distribution of $\tau$ (the variability in ORs in a population of treatment effects), elicited as before. We follow the approach by Chinn (2000) (37), where a continuous response is dichotomised, and a normal distribution is approximated by a logistic distribution.

A cut-off $c$ of interest is chosen, and the OR $\delta_i$ is defined as

$$\delta_i = \left(\frac{P(X_{i2} \geq c)}{P(X_{i2} < c)}\right) / \left(\frac{P(X_{i1} \geq c)}{P(X_{i1} < c)}\right). \qquad (1)$$

We can approximate a normal distribution $N(m, s^2)$ by a logistic distribution with same mean and variance, setting the location parameter in the logistic distribution to $m$ and the scale parameter to $\frac{s\sqrt{3}}{\pi}$. Using the logistic distribution approximation, the OR (1) is

$$\delta_i = \exp\left(\frac{\phi_i \pi}{\sqrt{3}}\right) = \exp\left(\frac{\beta_i \pi}{\sigma\sqrt{3}}\right).$$

We now have

$$\tau_{SMD} = \frac{\sqrt{3}\tau}{\pi},$$

$$\tau_{MD} = \frac{\sqrt{3}\sigma\tau}{\pi},$$

where $\tau$ is the between-study standard deviation (SD) on the log OR scale. Hence, we can now use the method in Section 3.1 with the following modification.

1. Dichotomise the response using some appropriate cut-off $c$, to define a new treatment effect $\delta_i$: the OR (1).

2. Considering ORs for the dichotomised response, use the three-stage procedure to elicit a prior distribution for $\tau$, the variability in ORs in a population of treatment effects.

3. Given a prior distribution for $\tau$, convert it to a prior distribution for the between-study SD $\tau_{MD}$ and $\tau_{SMD}$ on the continuous scale via $\tau_{MD} = \frac{\sqrt{3}\sigma\tau}{\pi}$ for MD, and $\tau_{SMD} = \frac{\sqrt{3}\tau}{\pi}$ for SMD, where $\sigma$ is an estimate of an individual level standard deviation. The estimate could be a summary measure of the SDs in the included studies, pooled from included studies, or obtained from a single representative study.

## 2      Elicitation method for ordered categorical data

For ordered categorical data, the likelihood function for the data would be a multinomial distribution with either a logit link function (i.e. a proportional odds model) or a probit link function. Suppose that there are $K$ outcome categories, denoted by $c_1, \dots, c_K$. Define $P_{ijk}$ to be the probability of an observation belonging to category $k$ or above, on treatment $j = 1,2$, with $j = 1$ the control arm and $j = 2$ the experimental treatment arm, in study $i$. For a logit link function, the treatment effect in the $i$th study can be defined by a single OR $\delta_i$, the OR

$$\frac{P_{i2k}}{1-P_{i2k}} \Big/ \frac{P_{i1k}}{1-P_{i1k}} \quad (2)$$

which is constant for all $k$. Hence, the outcome can be dichotomised into the two category sets $c_1, \dots, c_{k-1}$ and $c_k, \dots, c_K$, and the elicitation can proceed as in Section 3.1.

If a probit link function is used, the treatment effect in study $i$ may be described by a shift $\mu_i$ in the mean of the latent normal variable, and we again require a prior distribution for $\tilde{\tau}$, the variability in $\mu_i$ in a population of treatment effects. In this case, the OR (2) will change depending on the category $k$. However, an approximate prior for $\tau$ can be elicited using a similar approach to that in continuous outcome measures case: we dichotomise and approximate the latent normal variable by a latent logistic variable with scale parameter $\frac{\sqrt{3}}{\pi}$. We have the same modification as before:

1. Dichotomise the response using some appropriate category $c_k$, and define a new treatment effect $\delta_i$: the OR (2).

2. Use the three-stage procedure to elicit a prior distribution for $\tau$, the variability in ORs in a population of treatment effects.

3. Given a prior for $\tau$, convert this to a prior for $\tilde{\tau}$ via

$$\tilde{\tau} = \frac{\sqrt{3}}{\pi}\tau.$$

The interpretation of the heterogeneity parameter can be found in Table 1.

| Heterogeneity | 'range' of treatment effect, $R$, for scale-free outcome measure | $\tau$ for scale-free outcome measure | $\tau$ for outcome measure using probit or standardised mean difference scale | $\tau$ for outcome measure using mean difference scale |
|---|---|---|---|---|
| No heterogeneity | 1 | 0 | 0 | 0 |
| Low | 1.21 | 0.05 | 0.028 | $0.028\sigma$ |
| Moderate | 1.48 | 0.1 | 0.06 | $0.06\sigma$ |
|  | 2.19 | 0.2 | 0.11 | $0.1\sigma$ |
|  | 3.24 | 0.3 | 0.17 | $0.17\sigma$ |
|  | 4.80 | 0.4 | 0.22 | $0.22\sigma$ |
|  | 7.10 | 0.5 | 0.28 | $0.28\sigma$ |
| High | 10.51 | 0.6 | 0.33 | $0.33\sigma$ |
|  | 15.55 | 0.7 | 0.39 | $0.39\sigma$ |
|  | 23.01 | 0.8 | 0.44 | $0.44\sigma$ |
|  | 34.06 | 0.9 | 0.50 | $0.50\sigma$ |
|  | 50.40 | 1.0 | 0.55 | $0.55\sigma$ |
| Extremely high | 357.81 | 1.5 | 0.83 | $0.83\sigma$ |
|  | 2540.20 | 2 | 1.10 | $1.10\sigma$ |

Table 1: Suggested interpretation of the between-study standard deviation. The scale-free outcome measure refers to odds ratio, relative risk, hazard ratio and ratio of means. The estimate of $\sigma$ could be a summary measure of the standard deviations in the included studies, pooled from included studies, or obtained from a single representative study.

# Appendix 2

## 1      BUGS code for example TA163

```
model{                # *** PROGRAM STARTS
 for(i in 1:ns){           # LOOP THROUGH STUDIES
   w[i,1] <- 0          # adjustment for multi-arm trials is zero for control arm
   delta[i,1] <- 0         # treatment effect is zero for control arm
   mu[i] ~ dnorm(0,0.0001)       # vague priors for all trial baselines
   for (k in 1:na[i]) {         # LOOP THROUGH ARMS
     r[i,k] ~ dbin(p[i,k],n[i,k])   # binomial likelihood
     logit(p[i,k]) <- mu[i] + delta[i,k]  # model for linear predictor
     rhat[i,k] <- p[i,k] * n[i,k]   # expected value of the numerators
     dev[i,k] <- 2 * (r[i,k] * (log(r[i,k])-log(rhat[i,k]))
           + (n[i,k]-r[i,k]) * (log(n[i,k]-r[i,k]) - log(n[i,k]-rhat[i,k])))      #Deviance contribution  }
    resdev[i] <- sum(dev[i,1:na[i]])     # summed residual deviance contribution for this trial
   for (k in 2:na[i]) {         # LOOP THROUGH ARMS
    # trial-specific LOR distributions
    delta[i,k] ~ dnorm(md[i,k],precisiond[i,k])
    # mean of LOR distributions (with multi-arm trial correction)
    md[i,k] <-  d[t[i,k]] - d[t[i,1]] + sw[i,k]
    # precision of LOR distributions (with multi-arm trial correction)
    precisiond[i,k] <- precision *2*(k-1)/k
    # adjustment for multi-arm RCTs
    w[i,k] <- (delta[i,k] - d[t[i,k]] + d[t[i,1]])
    sw[i,k] <- sum(w[i,1:k-1])/(k-1)     # cumulative adjustment for multi-arm trials
   }
```

```
  }

  totresdev <- sum(resdev[])         # Total Residual Deviance

  d[1]<-0     # treatment effect is zero for reference treatment

  # vague priors for treatment effects

  for (k in 2:nt){  d[k] ~ dnorm(0,0.0001)

    OR[k] <- exp(d[k])

    d.new[k] ~ dnorm(d[k],precision)

    OR.new[k] <- exp(d.new[k])

  }

  # vague prior U[0,5] for between-trial SD

  tau ~ dunif(0,5)

  precision <- pow(tau,-2)   # between-trial precision = (1/between-trial variance)

  # informative prior using Turner et al (2012)

  #tau2~dlnorm(-2.57,0.33)   # prior for between study variance from lognormal (-2.57, 1.74^2)

  #tau<-sqrt(tau2)

  #precision<-1/tau2     # between-trial precision = (1/between-trial variance)

  # informative prior using Turner et al (2012) truncated so that the ratio of ORs can't exceed 10

  #tau2~dlnorm(-2.57,0.33)I(,0.345)    # R=exp(3.92tau)=> tau^2=(log(10)/3.92)^2=0.345

  #tau<-sqrt(tau2)

  #precision<-1/tau2      # between-trial precision = (1/between-trial variance)

  # informative prior using elicitation

  #R~dgamma(2.68,0.721) #elicited prior for the 'range' of OR

  #tau<-log(R+1)/3.92    #minimum of R is 1; convert the 'range' of OR to the between-study standard deviation

  #precision<-pow(tau,-2)      # between-trial precision = (1/between-trial variance)

}                     # *** PROGRAM ENDS
```

#Data (1=placebo, 2=infliximab, 3=ciclosporin)

list(ns=4,nt=3)

| t[,1] | t[,2] | n[,1] | r[,1] | n[,2] | r[,2] | na[] |
|---|---|---|---|---|---|---|
| 1 | 2 | 21 | 14 | 24 | 7 | 2 |
| 1 | 2 | 3 | 3 | 3 | 0 | 2 |
| 1 | 3 | 9 | 4 | 11 | 3 | 2 |
| 1 | 3 | 15 | 3 | 14 | 3 | 2 |

END

## 2 BUGS code for example TA336

```
model{                    # *** PROGRAM STARTS
 for(i in 1:ns){           #  LOOP THROUGH STUDIES
   w[i,1] <- 0            # adjustment for multi-arm trials is zero for control arm
   delta[i,1] <- 0        # treatment effect is zero for control arm
   mu[i] ~ dnorm(0,0.0001)     # vague priors for all trial baselines
   for (k in 1:na[i]) {      #  LOOP THROUGH ARMS
     var[i,k] <- pow(se[i,k],2)   # calculate variances
     prec[i,k] <- 1/var[i,k]     # set precisions
     y[i,k] ~ dnorm(theta[i,k],prec[i,k]) # binomial likelihood
     theta[i,k] <- mu[i] + delta[i,k]  # model for linear predictor
     dev[i,k] <- (y[i,k]-theta[i,k])*(y[i,k]-theta[i,k])*prec[i,k]  #Deviance contribution
   }
   # summed residual deviance contribution for this trial
   resdev[i] <- sum(dev[i,1:na[i]])
   for (k in 2:na[i]) {     # LOOP THROUGH ARMS
     # trial-specific distributions
```

```
      delta[i,k] ~ dnorm(md[i,k],precisiond[i,k])

      # mean of distributions, with multi-arm trial correction

      md[i,k] <-  d[t[i,k]] - d[t[i,1]] + sw[i,k]

      # precision of distributions (with multi-arm trial correction)

      precisiond[i,k] <- precision *2*(k-1)/k

      # adjustment, multi-arm RCTs

      w[i,k] <- (delta[i,k] - d[t[i,k]] + d[t[i,1]])

      # cumulative adjustment for multi-arm trials

      sw[i,k] <- sum(w[i,1:k-1])/(k-1)

  }

}

totresdev <- sum(resdev[])         #Total Residual Deviance

d[1]<-0      # treatment effect is zero for control arm

d.new[1]<-0

 # vague priors for treatment effects

for (k in 2:nt){  d[k] ~ dnorm(0,0.0001)

   d.new[k] ~ dnorm(d[k],precision) }

# vague prior U[0,5] for between-trial SD

tau ~ dunif(0,5)

precision <- pow(tau,-2)   # between-trial precision = (1/between-trial variance)

 # informative prior using Turner et al (2012) on odds ratio scale

#tau2~dlnorm(-2.56,0.33)  #odds ratio scale

#tau<-sqrt(tau2)/1.81*2.61   #mean difference scale; 2.61 is the mean of individual level standard deviation

#precision <- pow(tau,-2)   # between-trial precision = (1/between-trial variance)
```

# informative prior using Turner et al (2012) truncated so that the ratio of ORs can't exceed 10 on the odds ratio scale, R=exp(3.92tau)=>tau^2=(log(10)/3.92)^2=0.345

  #tau2~dlnorm(-2.56,0.33)I(,0.345)

  #tau<-sqrt(tau2)/1.81*2.61  #mean difference scale

  #precision <- pow(tau,-2)   # between-trial precision = (1/between-trial variance)

  # informative prior using elicitation

  #R~dgamma(1.94,0.823)        #odds ratio scale

  #tau<-log(R+1)/3.92/1.81*2.61  #mean difference scale, #minimum of R is 1

  #precision<-pow(tau,-2)       # between-trial precision = (1/between-trial variance)

}                     # *** PROGRAM ENDS

#Data (1=Placebo+Met+SU, 2=Sita+Met+SU, 3=Empa 10mg+Met+SU, 4=Lina+Met+SU, 5=Saxa+Met+SU, 6=Can 300mg+Met, 7=Can 100mg+Met, 8=Empa 25mg+Met+SU)

list(ns=6,nt=8)

| t[,1] | t[,2] | t[,3] | y[,1]  | y[,2]  | y[,3]  | se[,1] | se[,2] | se[,3] | na[] |
|-------|-------|-------|--------|--------|--------|--------|--------|--------|------|
| 1     | 3     | 8     | -0.39  | -2.16  | -2.39  | 0.15   | 0.15   | 0.16   | 3    |
| 1     | 4     | NA    | -0.06  | 0.27   | NA     | 0.16   | 0.09   | NA     | 2    |
| 1     | 2     | NA    | -0.70  | 0.40   | NA     | 0.3316 | 0.2551 | NA     | 2    |
| 1     | 5     | NA    | -0.60  | 0.20   | NA     | 0.1849 | 0.1945 | NA     | 2    |
| 2     | 6     | NA    | 0.2649 | -2.384 | NA     | 0.1325 | 0.1325 | NA     | 2    |
| 1     | 7     | 6     | -0.648 | -1.945 | -2.408 | 0.2362 | 0.2362 | 0.2362 | 3    |

END

# Appendix 3

## 1  R code instructions

A function `elicitHeterogen()` is available in the development version of the R package SHELF (38), which can be installed from GitHub with the commands

```
install.packages("devtools")
```

```
devtools::install_github("OakleyJ/SHELF")
```

The elicitation tool can then be run using the commands

```
library(SHELF)
```

```
elicitHeterogen()
```

Type `?elicitHeterogen` for further instructions.

## 2  Elicited prior distribution for the re-analysis of TA163 and TA336

Table 1 shows the number of bins used and the number of probs/chips allocated in each bin for the re-analysis of TA163 and TA336. The elicited prior for $R-1$ was gamma (2.62, 0.721). It presented the beliefs that the probability of heterogeneity being low, moderate and high as 0.01, 0.85, and 0.14, respectively. The R function used was `elicitHeterogen(lower=1,upper=10,nbins=9)`.

The elicited prior for $R-1$ was gamma (1.94, 0.741). It presented the beliefs that the probability of heterogeneity being low, moderate and high as 0.06, 0.88, and 0.06, respectively. The R function used was
`elicitHeterogen(lower=1,upper=10,nbins=9,sigma=2.61,scale.free=FALSE)`.

| Bin boundary | [1, 2) | [2, 3) | [3, 4) | [4, 5) | [5, 6) | [6, 7) | [7, 8) | [8,9) | [9, 10) |
|---|---|---|---|---|---|---|---|---|---|
| Number of probs allocated (TA136) | 4 | 5 | 6 | 6 | 5 | 4 | 2 | 1 | 1 |
| Number of probs allocated (TA336) | 4 | 5 | 4 | 3 | 2 | 1 | 1 | 0 | 0 |

Table 1: The number of bins and the number of probs allocated in each bin for the re-analysis of TA163 and TA336.